\documentclass[11pt]{article}

\usepackage{multirow}
\usepackage{subcaption}
\usepackage{graphicx}
\usepackage{authblk}
\usepackage{amsmath}
\usepackage{colortbl}
\usepackage[table]{xcolor}
\usepackage[margin=1in]{geometry}
\usepackage{float}

\definecolor{DartmouthGreen}{RGB}{0,105,62}

\begin{document}


\title{When minor issues matter: symmetries, pluralism, and polarization in similarity-based opinion dynamics}




\author[a,*]{Brian Mintz}
\author[b]{Daniel Simonson}
\author[c]{Dominik Wodarz}
\author[d]{Feng Fu}
\author[e,*]{Natalia L. Komarova}

\affil[a]{RIKEN Institute of Physical and Chemical Research, Wako, Japan} 
\affil[b]{Department of Mathematics, University of California Irvine, Irvine, CA 92697}
\affil[c]{Department of Ecology, Behavior, \& Evolution, University of California San Diego, 9500 Gilman Dr. La Jolla, CA 92093}
\affil[d]{Department of Mathematics, Dartmouth College, 29 N Main St, Hanover, NH 03755}
\affil[e]{Department of Mathematics, University of California San Diego, 9500 Gilman Dr. La Jolla, CA 92093}

\maketitle

\begin{abstract} 
Understanding how opinions evolve through social interactions is crucial for mitigating polarization. Existing opinion-dynamics models incorporate both attractive and repulsive interactions but typically assume that all issues are equally important. We develop and analyze a stochastic agent-based model where issues carry heterogeneous weights that influence both social affinity and the likelihood of opinion change. Surprisingly, introducing even a single issue with arbitrarily small weight can destabilize otherwise stable states, increasing convergence times by orders of magnitude. To explain these dynamics, we derive a mean-field approach and characterize the equilibrium symmetries governing consensus, polarization, and persistent pluralism. A complete classification of these symmetries for up to five issues reveals that polarization increases when importance is concentrated on a small number of issues. Conversely, distributing importance more broadly across issues promotes diversity of opinions and reduces polarization. Our symmetry-based framework highlights how issue salience and social tolerance jointly shape collective opinion evolution. 
\end{abstract}

Studying how individuals form and change their opinions through social interactions is fundamental to understanding cultural evolution, political polarization, and collective decision-making. Opinion dynamics models seek to capture these processes mathematically, revealing emergent patterns from local interactions that shape societal consensus or fragmentation \cite{levin2021dynamics,castellano2009statistical}. These models have found applications across diverse domains, from understanding the spread of political beliefs \cite{kawakatsu2021interindividual,wang2020public} to predicting market behaviors \cite{cont2000herd} and analyzing social media dynamics \cite{delvicario2016spreading, galam2024fake}.

Early foundational models established key frameworks for studying opinion dynamics. The voter model, where individuals randomly adopt neighbors' opinions, provided crucial insights into consensus formation through neutral drift processes \cite{clifford1973model,holley1975ergodic}. The DeGroot model introduced weighted influence networks, showing how opinion changes could be characterized through linear averaging processes \cite{degroot1974reaching}. The classic Deffuant–Weisbuch and  Hegselmann-Krause bounded confidence models incorporated the psychological principle that individuals only consider opinions sufficiently similar to their own \cite{deffuant2000mixing, hegselmann2002opinion}. These models revealed that smaller confidence bounds lead to more fragmented opinion clusters, capturing the echo chamber effect observed in real societies \cite{meng2018opinion}.

The recognition that opinions form across multiple interconnected issues led to important insights. Axelrod's pioneering cultural dissemination model demonstrated that homophily, i.e. the tendency to interact with similar others, could produce either global consensus or stable cultural diversity depending on the number of cultural features and their possible values \cite{axelrod1997dissemination}. Subsequent work has explored multidimensional opinion spaces with various geometric structures \cite{stamoulas2018convergence,etesami2013termination,baumann2021emergence}, revealing complex bifurcation patterns and phase transitions between consensus and fragmentation states. These studies connect to broader frameworks of cultural evolution and transmission developed by Cavalli-Sforza, Feldman, and colleagues \cite{cavallisforza1981cultural,feldman2017cultural}, which established mathematical foundations for understanding how cultural traits propagate through populations via diverse transmission mechanisms.




Recent work has emphasized the fundamental tension between individual interests and collective behavior across biological and social systems \cite{levin2010crossing,levin2014public}. It has been recognized that social interactions involve not only attraction (homophily with similar others) but also repulsion (differentiation from dissimilar others). While a number of different outcomes have been reported in models with attractive interactions \cite{pedraza2021analytical, cheng2025multidimensional}, models incorporating both forces reveal richer dynamics. For instance, antagonistic interactions can drive polarization where populations split into opposing camps  while the balance between attraction and repulsion determines whether systems reach consensus, polarization, pluralism, or other states \cite{kurmyshev2011dynamics, balenzuela2015undecided, chen2017deffuant, cui2023exploring,huang2024breaking, lanchier2024deffuant}.  Understanding these dynamics within and among competing groups has become central to explaining social phenomena from cooperation to political polarization \cite{sabin2020pull,stewart2021inequality, cooney2023evolutionary,levin2023polarization}. Network structure further modulates these effects, with heterogeneous connectivity patterns influencing convergence times and equilibrium states \cite{sood2005voter,parsegov2017novel}. The evolution of cultural traits under frequency-dependent selection adds further richness to opinion dynamics \cite{newberry2022measuring,hirshleifer2021moonshots}, showing how the prevalence of opinions can influence their tendency to spread. Moreover, collective intelligence emerges from information exchange and coordination in the service of public goods \cite{leonard2022collective}, yet these same mechanisms can be undermined by asymmetries in social interactions and information flow \cite{su2021evolution}.\\

Despite these advances, two critical aspects remain understudied. First, most models treat all issues as equally important, yet in reality, people weight different topics very differently; for example, taxation policy may matter more to people compared with, say, sports preferences. Second, the set of salient issues itself evolves: new topics emerge in public discourse while others fade. \\

Evidence of differential issue importance for people is abundant in the literature, see e.g. \cite{converse2006nature, krosnick1990government}.  Furthermore, it has been shown that issue importance has significant behavioral consequences for people:   individuals who regard the issue as highly important tend to use it much more heavily when evaluating candidates or choosing how to vote \cite{krosnick1988role}. In SI Section 1 we present some quantitative data from the literature showing empirical evidence of different opinion importance among US individuals. Related to this is the phenomenon of the rise of new salient issues, e.g. due to media coverage \cite{mccombs1972agenda}. It has been empirically shown that some issues may remain  important for extended time-periods, while others can fade away, and new ones emerge suddenly \cite{carmines1989issue}. Ref \cite{baumgartner2010agendas}  discussed this in the context of the punctuated equilibrium theory in American policymaking, arguing that long periods of political stability are interrupted by sudden, dramatic shifts in public policy agendas.\\

The present work extends the opinion dynamics modeling framework by including heterogeneous issue importance. We formulate a model that incorporates issue weights and study how people's opinions evolve when they hold views on multiple issues of varying importance, and when interactions can be either friendly (leading people to agree more) or antagonistic (pushing opinions apart). Opinion weights may play a dual role in our model: (1) they define how much a person's opinion on a given issue counts towards a ``friendship score", that is, the likelihood of an individual to engage in friendly interactions with another person; and (2) opinions on more important issues may be more stable (less likely to change) in individuals. We find that when people require high similarity to be ``friends," populations typically reach extreme polarization or persistent pluralism, while lower similarity thresholds promote consensus. 
Surprisingly, introducing even a single new issue of trivial importance can dramatically change these dynamics, destabilizing stable consensus or polarization states, and sometimes causing convergence times to increase by orders of magnitude. We provide a complete theoretical characterization explaining when and why these counterintuitive effects occur, based on underlying symmetries in the system. Our analysis reveals that the balance between friendly and antagonistic interactions, combined with how people weigh different issues, creates tipping points where small changes can trigger major shifts between consensus, polarization, and persistent diversity. These findings have important implications for understanding how discourse on new topics (even seemingly minor ones) can fundamentally reshape opinion landscapes in populations. \\

We first present a stochastic agent-based model with heterogeneous issue weights and derive the corresponding deterministic system in the large population limit. We then report numerical results showing how convergence behavior and final states depend on the similarity threshold and weight configurations. Through systematic analysis, we classify equilibrium symmetries and explain how these symmetries account for the observed dynamics. We conclude by discussing implications for understanding real-world opinion dynamics.

\section*{Model}

\paragraph{Stochastic model of opinion dynamics.} We consider a stochastic, agent-based model with a finite population of $N$ individuals, each having an independent binary string of opinions on $L$ different issues. Each issue $i$ has a positive weight $w_i$ representing its relative importance, and we assume (without loss of generality) that the list of weights is non-increasing and $\sum_{i=1}^L w_i=1$. For two binary opinion  strings (with length $L$), $s$ and $q$, we define their similarity, $\sigma(s,q)$, as
\begin{equation}
    \sigma(s,q) = \sum_{i=1}^L w_i \delta(s_i,q_i),
\end{equation}
where $\delta(x,y)$ is one if $x=y$ and zero otherwise.   That is, the similarity is the sum of the weights of issues where the pair agrees. In this  version of the model we assume that the weights are shared among the individuals. In SI Section 11 we  extend this framework to a heterogeneous setting.

\begin{figure*} 
    \centering
    \includegraphics[width=0.9\linewidth]{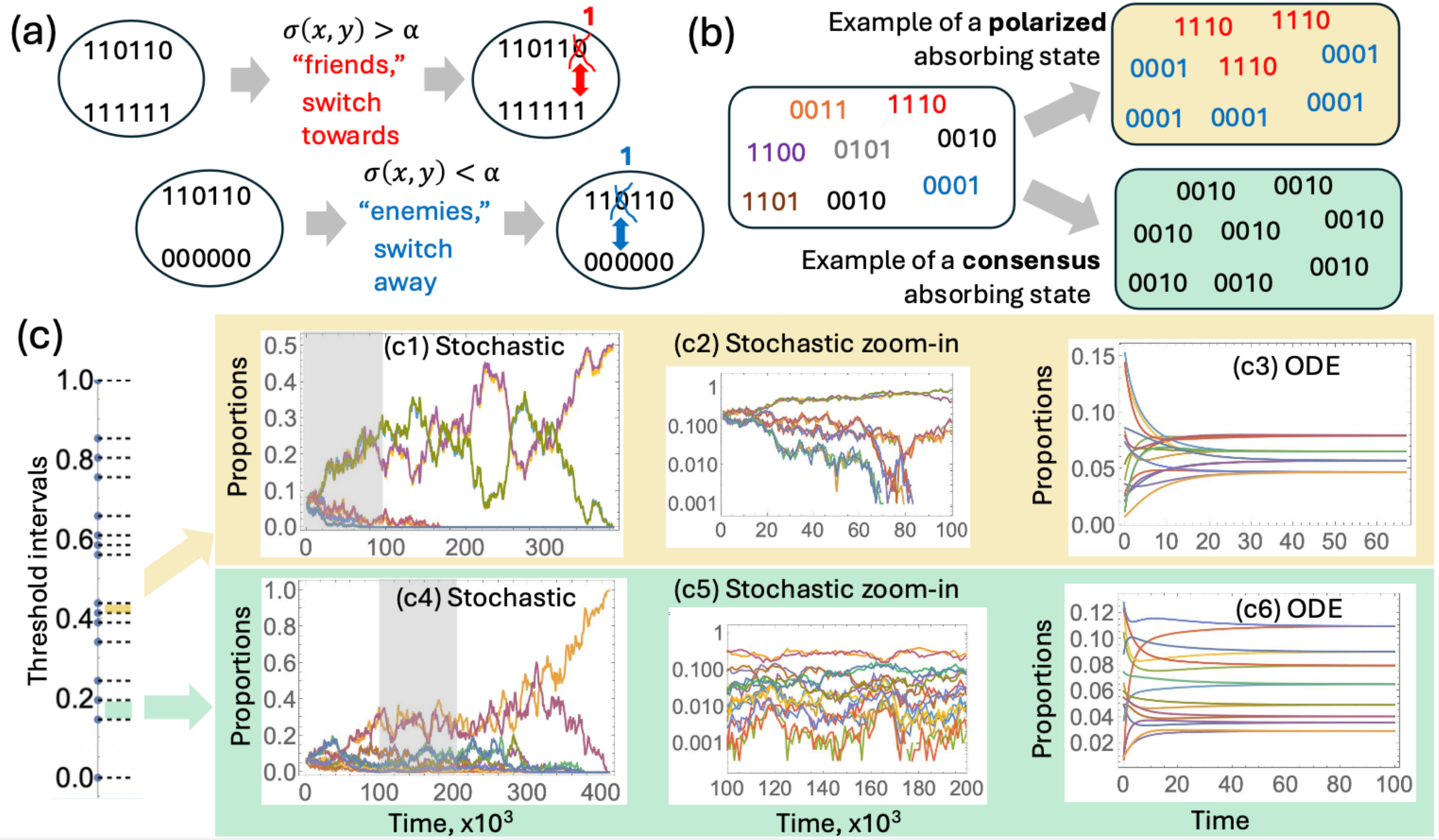}
    \caption{\textbf{Stochastic and deterministic dynamics of the opinion model}. (a) The rules of opinion change are illustrated using two interactions in an $L=6$ system, where in the first case, the similarity happens to be higher than the threshold, $\alpha$, and the focal individual (the upper string) updates one of the issues to be more like the interlocutor (the string lower string); in the second case, the similarity is lower than the threshold and the focal individual updates one of the issues to be further away from the interlocutor. (b) Two types of absorbing states of the stochastic model with $r=0$ are illustrated with an $L=4$ example. (c) Examples of an $L=4$ system behavior for two different threshold values. The intervals of $\alpha$ are shown on the left for weights $\frac{1}{205}[85,50,40,30]$ (case 8 of Table~\ref{tab:symmetries}). For intervals 2 (panels (c4,c5,c6)) and 7 (panels (c1,c2,c3)), a typical stochastic trajectory (left, (c1) and (c4)) is shown, together with a zoomed-in grayed fragment on a log-scale (middle, (c2) and (c5)), and a typical solution of the corresponding ODE (right, (c3) and (c6)). Different lines correspond to relative abundance of each of the 16 possible opinion strings. $N=3,000$ individuals, $r=0$, $\gamma_i=1$.}
    \label{fig:mod}
\end{figure*}

The dynamics are modeled as discrete updates. At each iteration, a focal and target individual are chosen uniformly at random to interact, discussing one of the $L$ issues, also chosen uniformly at random. With small probability $r$, the focal individual flips their opinion on the issue discussed, representing noise or the rare possibility of spontaneous innovations in opinion strings. Otherwise, for the majority of iterations, that is with probability $1-r$, the focal individual updates their opinion on the chosen issue based on that of the other individual and the degree of similarity between the two individuals, see illustration in figure \ref{fig:mod}(a). If this similarity is below a lower threshold $\alpha_e$, then the opinion strings are sufficiently different that the individuals are ``enemies", so the focal individual adopts the opposite  opinion of the interlocutor on the issue discussed. If instead the similarity is above an upper threshold $\alpha_f$, then the individuals are similar enough to be ``friends", and the focal individual updates their opinion on the chosen issue to match that of the target individual. If neither inequality is satisfied, the focal individual does not change their opinion on the issue discussed. These effects can be seen as repulsion and attraction respectively, with the parameters $\alpha_e$ and $\alpha_f$ governing the balance between these forces. The likelihood of change of an opinion might be a function of the issue's weight. We incorporate this by assuming that the rate of change of opinion on issue $i$ is $\gamma_i$, which may decay with $w_i$. A total of $N$ such updates (by the number of individuals) comprises a time-unit for this model.

\paragraph{Deterministic systems of opinion evolution.}
To better understand the dynamics of the stochastic model, we take a large population limit to obtain a deterministic model of a system of ordinary differential equations. Letting $x_s$ represent the proportion of the population with opinion string $s$ and averaging over all possible interactions, the dynamics with no noise ($r=0$) are given by equation (\ref{master}):
\begin{eqnarray}
\frac{d}{dt} x_{s}&=&\frac{1}{L}\sum_{i=1}^L\gamma_i\left(-x_s\left[\sum_{\sigma(s,q)>\alpha_f}x_q(1-\delta(s_i,q_i))+\sum_{\sigma(s,q)<\alpha_e}x_q\delta(s_i,q_i)\right]\right.\nonumber\\
\label{master}
&+&\left.\sum_{d(s,q)=1}x_q (1-\delta(s_i,q_i))\left[\sum_{\sigma(q,v)>\alpha_f}x_v\delta(s_i,v_i)+\sum_{\sigma(q,v)<\alpha_e}x_v(1-\delta(s_i,v_i))\right]\right).
\end{eqnarray}
The right hand side sums over all the issues, and the rate of change denoted by $\gamma_i$ is a function of the issues' weight. In particular, one could assume that $\gamma_i$ is a decaying function of $w_i$. The first term on the right describes decreases  in the proportion of opinion string $s$ in the population.  The frequency of string $s$ can decrease because (1) an individual with opinion string $s$  discusses one of the  issues that they disagree on with a ``friend" of opinion string $q$, or (2) an individual with opinion string $s$  discusses one of the  issues they agree on with an ``enemy" $q$. 

The second term on the right describes increases  in the proportion of opinion string $s$. Denote by $d(s,q)$ the Hamming distance between  opinion strings $s$ and $q$, that is, the number of issues on which they differ. We will refer to opinions strings $q$ with $d(s,q)=1$ as ``neighboring" opinion strings with respect to string $s$ (that is, they only differ from $s$ by a single issue). Increases in $x_s$  come from one of its neighboring types, $q$ ($d(s,q)=1$), flipping its opinion on issue $i$ such that the resulting string becomes $s$. This can come from two mechanisms: (1) A neighboring opinion string, $q$, interacts with its ``friend", $v$, whose opinion on $i$ is the same as the focal individual's ($s$). Then the neighboring string $q$ switches opinion $i$ to agree  with  ``friend" $v$, and as a result it is now identical to $s$. (2) A neighboring opinion string, $q$, interacts with its ``enemy", $v$, whose opinion on $i$ is different from the focal individual's ($s$). The neighboring string $q$ then switches opinion $i$ to disagree  with  ``enemy" $v$, and as a result it is now identical to $s$. 

For the majority of this work, we assume that $\gamma_i=1$, that is, the issue's weight does not influence the rate at which it changes. An extension of the model where $\gamma_i$ decreases with $w_i$ is studied in SI Section 10. Further, we will mostly take $\alpha\equiv \alpha_f = \alpha_e$ for simplicity, omitting the subscripts and simply writing $\alpha$ when this is the case (see SI Section 9 for a study of the case $\alpha_e \ne \alpha_f$).
Note that as parameter $\alpha$ is varied, the functional form of the right hand side of equation (\ref{master}) changes, see, e.g., Figure S2 of the SI. 

\section*{Results}

We seek to understand these large systems of $2^L$ variables with respect to $L$ parameters, which are the normalized weights $w_i$, and the similarity threshold $\alpha$. 
While the similarity threshold can be varied continuously from $0$ to $1$, the system behavior will change as a sequence of transitions, as $\alpha$ crosses each of $2^L$ threshold values. The reason for this is as follows. For a given set of weights, there are only finitely many values for the similarity of two opinion strings, as these are sums of one of the $2^L$ many subsets of the weights. Denote these threshold values as $0=\sigma_0,\sigma_1,\ldots,\sigma_{2^L}=1$. An example is presented in figure \ref{fig:mod}(c), left, where for an $L=4$ system, all the 16 values are shown as tick marks on $[0,1]$. These values split the domain of $\alpha$ into 15 intervals, and the behavior of the system (that is, its microscopic rules) only depends on which of these intervals, $(\sigma_{j-1},\sigma_j)$, $\alpha$ falls in, and not on the specific value of $\alpha$ within the interval. Two examples (for intervals 3 and 7 of this specific system) are shown, both for the stochastic and deterministic models.

\paragraph{Stochastic agent-based simulations.}
In stochastic simulations of opinion dynamics in our model in the absence of noise ($r=0$), only two types of absorbing states are possible (see Fig.~\ref{fig:mod}(b)): (i) the consensus state where every individual in the population has exactly the same opinions (see also example of  panel (c4) in Fig.~\ref{fig:mod}(c), where only a single string type remains), or (ii) the polarization state where the population is split into two groups; the opinions within each group are identical, and they are the opposite between the groups (see also example of  panel (c1) in Fig.~\ref{fig:mod}(c), where two string types remain, depicted by purple and yellow lines, both reaching approximately 50\%). Any other configuration will contain ``friends'' who differ on some issue or ``enemies'' who agree on some issue, allowing for further change. When $r \neq 0$, the population does not get stuck in these states, however these remain meaningful to study as they become long-lived states. Regardless of noise,  apart from the two types of absorbing states,  there is also a third type of outcome, which is a long-lived quasi-equilibrium state, which can be described as ``persistent pluralism", where all the possible opinion strings are equally represented in the population.

Using stochastic simulations, we asked: (1) What is the probability, under different parameter values, that the population converges to a consensus or a polarization state,  and (2) how long does convergence to either of the two absorbing states take? 

\begin{figure*}
    \centering
    \includegraphics[width=\linewidth]{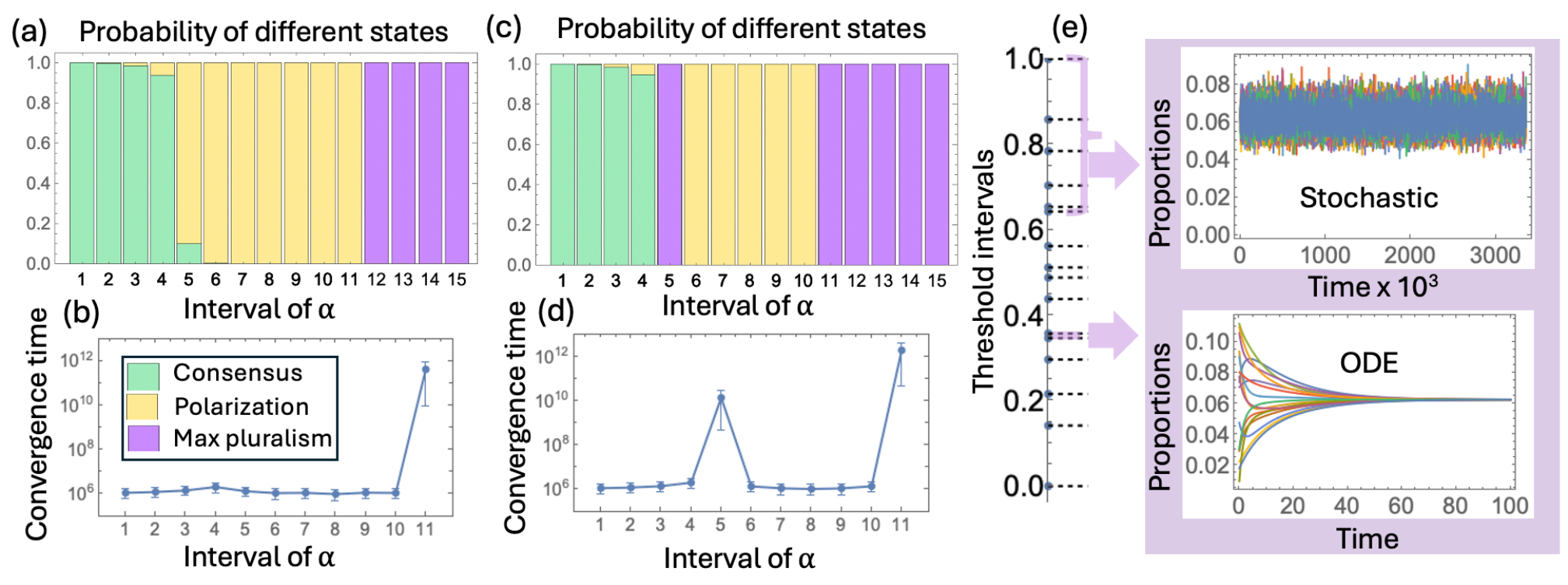}
    \caption{\textbf{Convergence behavior of the stochastic model varies significantly with $\alpha$.}  (a) The effect of threshold, $\alpha$, on the probability of the stochastic model ending in a consensus/polarization/persistent pluralism state (green/yellow/purple, respectively), when initialized uniformly at random, where the length of the bars represents the probability of different outcomes for different $\alpha$ intervals. Parameters are $L=4$, $N=100$, $\gamma_i=1$, $r=0$, and weights $\frac{1}{205}[85, 50, 40, 30]$ (case 8 of Table~\ref{tab:symmetries}). (b) The effect of $\alpha$ on the absorption times with the same weights as in (a), and initial conditions chosen uniformly at random. The inset explains the color code for panels (a) and (c). (c,d) The same as (a,b), except the weights are $[0.346494 , 0.29631 , 0.215498 , 0.141697]$ (case 13 of Table~\ref{tab:symmetries}). The number of simulations in (a-c) is $2,000$ per $\alpha$, except for the largest intervals with persistent pluralism, where we used $100$ simulations.  Other parameters are  $L=4$, $N=100$, and $r=0$. (e)  The intervals of $\alpha$ for the weights in (c,d) are shown on the left. For intervals 5 and 11-15, a typical stochastic trajectory  and the ODE solution are shown. Different lines correspond to relative abundance of each of the 16 possible opinion strings. Note the results for $\alpha$ intervals 12-15 are omitted in panels (b) and (d) for the clarity of the figure, as they were significantly larger than the other values.}
    \label{fig:effects}
\end{figure*}

We observe that the probability of consensus shows a clear decrease with $\alpha$, see the green bars in Fig. \ref{fig:effects} (a). Here results are presented for a specific choice of weights with $L=4$, but as we will see, this pattern holds more generally (see also Sections 5 and 6 of SI).  Larger values of threshold $\alpha$ correspond to more antagonistic interactions, preventing consensus and often resulting in a polarized state (yellow bars). 
The largest values of $\alpha$ lead to a state of persistent pluralism (purple bars). Note that the horizontal axis in this and other panels represents intervals of $\alpha$, as the microscopic rules stay the same for all the $\alpha$ values within those intervals.

Fig. \ref{fig:effects} (b) shows a typical dependence of the system convergence time on the threshold $\alpha$. For the same set of weights as in panel (a), we obtained a numerical distribution for the time to convergence. An increase in $\alpha$ beyond a certain value leads to a significant increase in convergence times. In fact, as the population increases, time to convergence becomes prohibitively long. The population settles in a state of persistent pluralism, where all the opinion strings are present at roughly equal proportions.   Intuitively, this result says that when $\alpha$ becomes large, where individuals require a high level of similarity to be ``friends", the population takes a long time to eventually polarize into opposing groups. In contrast, consensus is achieved much quicker when friendly interactions are common.

Surprisingly, for some specific combinations of opinion weights, a somewhat less regular pattern is observed, see panels (c,d) of Fig. \ref{fig:effects}. The only difference between panels (a,b) and (c,d) is a choice of weights, which results in  a different dependence on the similarity threshold $\alpha$, panel (d). For this case, a small change in $\alpha$ can have a very large effect on the model behavior. This only occurs for some weight configurations, as described in the supplement section 8. For $L=4$ issues, this type of non-monotonic behavior  requires the weights $[a, b, c, d]$ to satisfy $a<\alpha<d+c$. We also observed that this type of pattern becomes  more common for larger numbers of issues $L$ (it  never occurs for  $L=2$ or $L=3$). As we will show below, this strongly non-monotonic behavior is a consequence of inherent symmetries of the system, and can give rise to counterintuitive dynamics, as we show next.

\begin{figure*} 
    \centering
    \includegraphics[width=\linewidth]{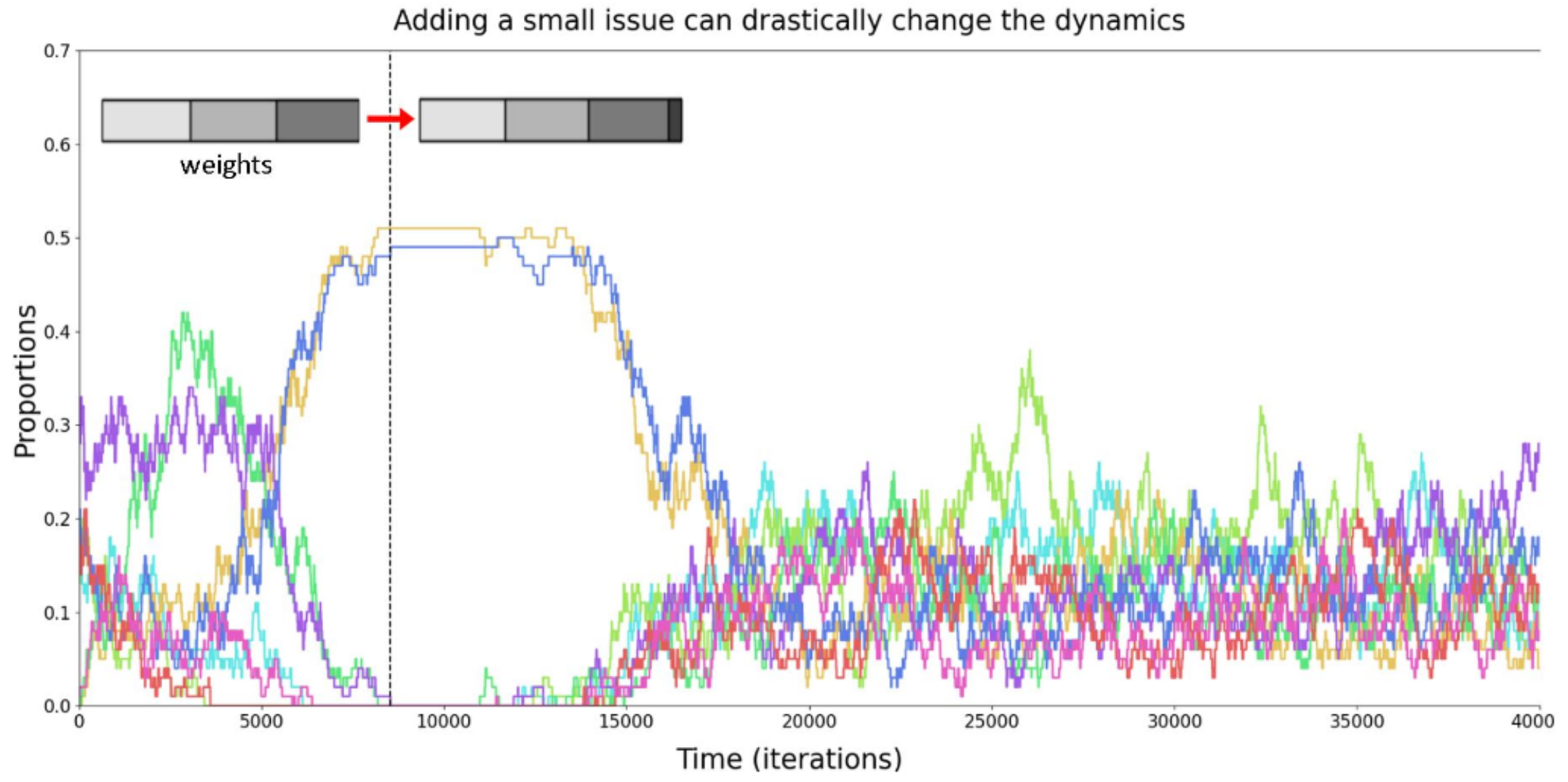}
    
    \caption{\textbf{Introducing issues of small weight can destabilize a population, causing every opinion string being held equally and significantly delaying convergence}. The proportions of each opinion string in a run of the stochastic model with weights $\frac{1}{3003}[1002, 1001, 1000]\approx[0.3336,0.3333,0.3331]$, $\alpha=\frac{1003}{3003}\approx0.334$, $N=100$, $\gamma_i=1$, and $r=0.001$. Once the population converges, in this case to a polarization state a bit before ten thousand iterations, a fourth issue is added (i.e. the weights are changed to $\frac{1}{3008}[1002, 1001, 1000,5]\approx[0.3331, 0.3328, 0.3324, 0.0017]$) and individuals with opinion string $s$ are split uniformly at random into the opinion strings $s0$ and $s1$. For clearer comparison of the dynamics, opinion strings with the same opinions on the first three issues are combined for plotting. The choice of these weights is explained in the supplement, and follows the symmetries we determined. Indeed, we observe the predicted symmetries where polar opposites have equal proportions (corresponding to the third row and fifth column of Fig. S3), and then to $C_0$ in case thirteen and $\alpha$ interval five of \ref{tab:symmetries}, a state of persistent pluralism. }
    \label{fig:add-issue-and-symmetry}
\end{figure*}

Fig. \ref{fig:add-issue-and-symmetry} presents a simulation where, in the presence of noise, adding a single issue of arbitrarily small weight can destabilize a population that has polarized. The set up is as follows. We first run a simulation with a fixed set of parameters, then, once consensus or polarization is reached, we leave $\alpha$ unchanged but add a new issue of small weight. In most cases, the population would quickly reach convergence again, often to the same type of state, that is, a polarized population will polarize again, and a population in consensus will find consensus after this disturbance. However, as we see in this figure, it is possible that introducing a low-weight issue can lead to a destabilization of the equilibrium, and eventually results in a state of persistent pluralism, see SI Section 8 for more details of this type of  simulations.

The fact that a single  issue with a small weight can have such large effects seems counterintuitive, but follows from symmetries of the system. To observe this effect, two conditions must hold: (i) the system with the additional issue(s) must have a super-symmetric solution, as explained below, and (ii) there is sufficient diversity in the opinion strings in a population when an additional issue is introduced. When there is no noise, and the population has reached an absorbing state, adding a single issue leads to only a few opinions strings being present in the population, resulting in quick polarization or consensus. Noise can create enough diversity in the opinion strings (which is what we see in figure \ref{fig:add-issue-and-symmetry}. The same effect occurs in the noiseless model if the issue is introduced before the population reaches consensus or polarization, or if more than one issue is added (though we did not identify a case were the combined weight of the two issues could be made arbitrarily small). This is explained further in the subsection describing the neutral drift manifold.
Intriguingly, we found no cases where the addition of an issue made polarization more likely, see SI Section 8.

\paragraph{Abundance of symmetry in equilibria.}
To begin understanding these effects, we turn to the trajectories in the deterministic system of differential equations with no noise ($r=0$). It turns out that depending on the $\alpha$ interval, the structure of the equilibrium solution changes. Panel (c) of Fig.~\ref{fig:mod} shows two examples of the $\alpha$ threshold, for the same system of weights in an $L=4$ case. The bottom row (panels (c4,c5,c6)) uses an $\alpha$ value from the 2nd interval. The stochastic trajectory (c4) results in a consensus absorbing state (expected for relatively small values of $\alpha$, see figure  \ref{fig:effects}(a)), but when we zoom in the gray area of panel (c4) (see panel (c5)), we can see that for part of the time, different opinion  strings are arranged in pairs. In this simulation, the abundances of strings 0000 and 0001, 0010 and 0011, 0100 and 1011, etc appear highly correlated with each other. To express the general pattern of this symmetry, we will use the notations $xyz*$. Here, the symbols $x$, $y$, and $z$ refer to a specific choice of an opinion, and symbol $*$ means ``any". So, for each choice of a binary string, $xyz$, we have a set of two strings, $xyz0$ and $xyz1$, whose abundances are  highly correlated. Figure \ref{fig:mod} panel (c6) shows a simulation of the corresponding ODE, and we can see that at the equilibrium, the different populations again arrange in pairs of the same abundance. Computation of the eigenvalues of the Jacobian at these points shows that this steady state is neutrally stable, pointing at the existence of a stable manifold, where the abundances are equal within each pair, but their values  are arbitrary.

The second example considers the seventh $\alpha$ interval for the same system of issue weights. The stochastic run shown in panel (c1) of figure \ref{fig:mod}(c) results in a polarized state (as expected for this interval of $\alpha$, see figure  \ref{fig:effects}(a)), where two populations of polar opposite opinion strings end up having similar abundances. Zooming in (c2) we can see that initially, the 8 opinion strings aggregate in groups of four populations, such as for each choice of a binary string $xyz$,  strings $xyz*$ and $\bar x\bar y \bar z *$  comprise a group, where the bar denotes the polar opposite. The same type of symmetry is observed in the neutrally stable steady state of the corresponding ODE (panel (c3)), where the 16 populations split into four groups that take the same values, although the values themselves will depend on the initial condition. 

\begin{figure*} 
    \centering
    \includegraphics[width=\linewidth]{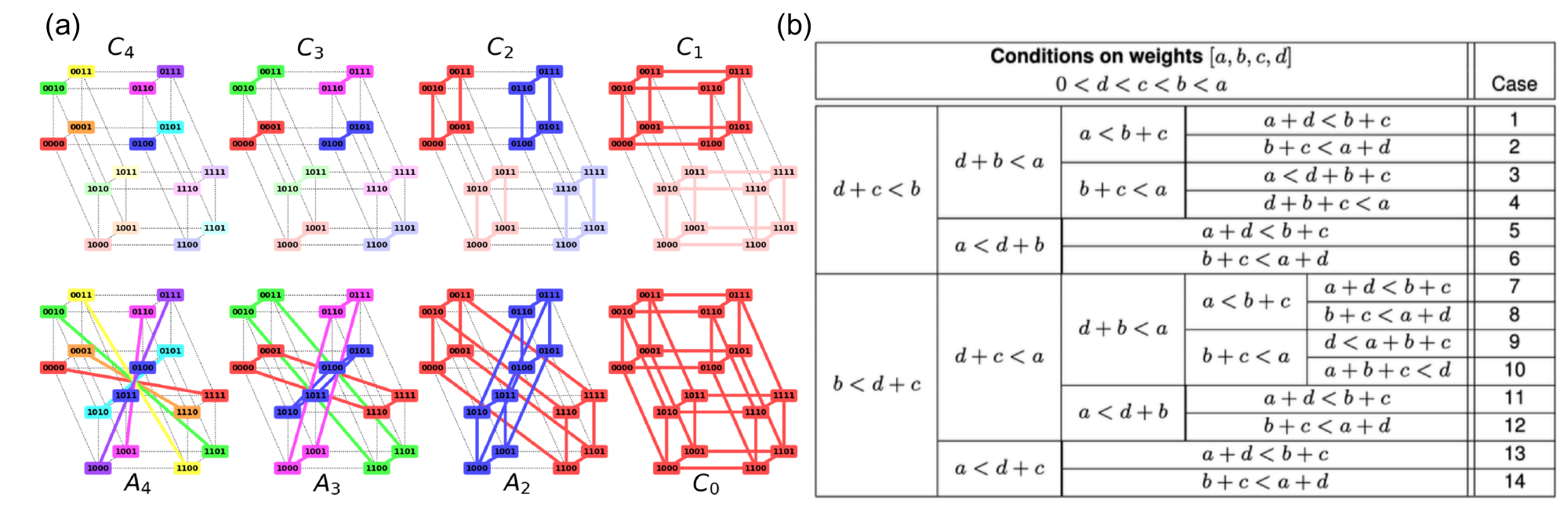}
    \caption{\textbf{Classification of symmetries on the deterministic model with $L=4$.} (a) Visual depiction of symmetries. The diagrams  represent the possible symmetries seen among frequencies of opinion strings at equilibria in the deterministic model. Dashed lines are used to show the  of the hypercube $\{0,1\}^4$ for reference. Solid lines connect opinion strings with equal proportions at equilibrium, along with matching colors. The lighter shades have no relation to the more saturated colors, and are just used as a convenient set of additional colors. (b) The 14 cases of weights. A complete characterization of when each symmetry occurs for each of these cases is given in Table~\ref{tab:symmetries}.} 
    \label{fig:symmetries}
\end{figure*}

\vspace{3cm}

\begin{table}[H]
    \resizebox{\textwidth}{!}{
         \begin{tabular}{|c||c|c|c|c|c|c|c|c|c|c|c|c|c|c|c|}
        \hline
            &  \multicolumn{15}{c|}{\textbf{Interval for Similarity threshold} $\alpha$} \\
        \hline
        Case & 1 & 2 & 3 & 4 & 5 & 6 & 7 & 8 & 9 & 10 & 11 & 12 & 13 & 14 & 15 \\
        \hline
        \hline
             1 & \cellcolor{green!20}$C_4$ & \cellcolor{green!20}$C_3$ & \cellcolor{green!20}$C_2$ & \cellcolor{yellow!20}$C_2$ & \cellcolor{yellow!20}$C_2$ & \cellcolor{yellow!20}$C_3$ & \cellcolor{red!20}$A_3$ &  \cellcolor{red!20}$A_3$ &  \cellcolor{red!20}$A_3$ &  \cellcolor{red!20}$A_3$ &  \cellcolor{red!20}$A_2$ &  \cellcolor{red!20}$A_2$ &  \cellcolor{violet!30}$C_0$   & \cellcolor{violet!30}$C_0$   & \cellcolor{violet!30}$C_0$    \\
        \hline
             2  & \cellcolor{green!20}$C_4$ & \cellcolor{green!20}$C_3$ & \cellcolor{green!20}$C_2$ & \cellcolor{yellow!20}$C_2$ & \cellcolor{yellow!20}$C_2$ & \cellcolor{yellow!20}$C_3$ & \cellcolor{red!20}$A_3$ &  \cellcolor{red!20}$A_4$ &  \cellcolor{red!20}$A_3$ &  \cellcolor{red!20}$A_3$ &  \cellcolor{red!20}$A_2$ &  \cellcolor{red!20}$A_2$ &  \cellcolor{violet!30}$C_0$   & \cellcolor{violet!30}$C_0$   & \cellcolor{violet!30}$C_0$    \\
        \hline
             3  & \cellcolor{green!20}$C_4$ & \cellcolor{green!20}$C_3$ & \cellcolor{green!20}$C_2$ & \cellcolor{yellow!20}$C_2$ & \cellcolor{yellow!20}$C_2$ & \cellcolor{yellow!20}$C_3$ & \cellcolor{red!20}$C_4$ & \cellcolor{red!20}$A_4$ &  \cellcolor{red!20}$A_4$ &  \cellcolor{red!20}$A_3$ &  \cellcolor{red!20}$A_2$ &  \cellcolor{red!20}$A_2$ &  \cellcolor{violet!30}$C_0$   & \cellcolor{violet!30}$C_0$   & \cellcolor{violet!30}$C_0$    \\
        \hline
             4  & \cellcolor{green!20}$C_4$ & \cellcolor{green!20}$C_3$ & \cellcolor{green!20}$C_2$ & \cellcolor{yellow!20}$C_2$ & \cellcolor{yellow!20}$C_2$ & \cellcolor{yellow!20}$C_3$ & \cellcolor{red!20}$C_4$ & \cellcolor{red!20}$C_4$ & \cellcolor{red!20}$A_4$ &  \cellcolor{red!20}$A_3$ &  \cellcolor{red!20}$A_2$ &  \cellcolor{red!20}$A_2$ &  \cellcolor{violet!30}$C_0$   & \cellcolor{violet!30}$C_0$   & \cellcolor{violet!30}$C_0$    \\
        \hline
             5  & \cellcolor{green!20}$C_4$ & \cellcolor{green!20}$C_3$ & \cellcolor{green!20}$C_2$ & \cellcolor{yellow!20}$C_2$ & \cellcolor{yellow!20}$C_2$ & \cellcolor{red!20}$A_2$ &  \cellcolor{red!20}$A_3$ &  \cellcolor{red!20}$A_3$ &  \cellcolor{red!20}$A_3$ &  \cellcolor{red!20}$A_2$ &  \cellcolor{red!20}$A_2$ &  \cellcolor{red!20}$A_2$ &  \cellcolor{violet!30}$C_0$   & \cellcolor{violet!30}$C_0$   & \cellcolor{violet!30}$C_0$    \\
        \hline
             6  & \cellcolor{green!20}$C_4$ & \cellcolor{green!20}$C_3$ & \cellcolor{green!20}$C_2$ & \cellcolor{yellow!20}$C_2$ & \cellcolor{yellow!20}$C_2$ & \cellcolor{red!20}$A_2$ &  \cellcolor{red!20}$A_3$ &  \cellcolor{red!20}$A_4$ &  \cellcolor{red!20}$A_3$ &  \cellcolor{red!20}$A_2$ &  \cellcolor{red!20}$A_2$ &  \cellcolor{red!20}$A_2$ &  \cellcolor{violet!30}$C_0$   & \cellcolor{violet!30}$C_0$   & \cellcolor{violet!30}$C_0$    \\
        \hline
             7  & \cellcolor{green!20}$C_4$ & \cellcolor{green!20}$C_3$ & \cellcolor{green!20}$C_2$ & \cellcolor{green!20}$C_1$ & \cellcolor{yellow!20}$C_2$ & \cellcolor{yellow!20}$C_3$ & \cellcolor{red!20}$A_3$ &  \cellcolor{red!20}$A_3$ &  \cellcolor{red!20}$A_3$ &  \cellcolor{red!20}$A_3$ &  \cellcolor{red!20}$A_2$ &  \cellcolor{violet!30}$C_0$   & \cellcolor{violet!30}$C_0$   & \cellcolor{violet!30}$C_0$   & \cellcolor{violet!30}$C_0$    \\
        \hline
             8  & \cellcolor{green!20}$C_4$ & \cellcolor{green!20}$C_3$ & \cellcolor{green!20}$C_2$ & \cellcolor{green!20}$C_1$ & \cellcolor{yellow!20}$C_2$ & \cellcolor{yellow!20}$C_3$ & \cellcolor{red!20}$A_3$ &  \cellcolor{red!20}$A_4$ &  \cellcolor{red!20}$A_3$ &  \cellcolor{red!20}$A_3$ &  \cellcolor{red!20}$A_2$ &  \cellcolor{violet!30}$C_0$   & \cellcolor{violet!30}$C_0$   & \cellcolor{violet!30}$C_0$   & \cellcolor{violet!30}$C_0$    \\
        \hline
             9  & \cellcolor{green!20}$C_4$ & \cellcolor{green!20}$C_3$ & \cellcolor{green!20}$C_2$ & \cellcolor{green!20}$C_1$ & \cellcolor{yellow!20}$C_2$ & \cellcolor{yellow!20}$C_3$ & \cellcolor{red!20}$C_4$ & \cellcolor{red!20}$A_4$ &  \cellcolor{red!20}$A_4$ &  \cellcolor{red!20}$A_3$ &  \cellcolor{red!20}$A_2$ &  \cellcolor{violet!30}$C_0$   & \cellcolor{violet!30}$C_0$   & \cellcolor{violet!30}$C_0$   & \cellcolor{violet!30}$C_0$    \\
        \hline
             10  & \cellcolor{green!20}$C_4$ & \cellcolor{green!20}$C_3$ & \cellcolor{green!20}$C_2$ & \cellcolor{green!20}$C_1$ & \cellcolor{yellow!20}$C_2$ & \cellcolor{yellow!20}$C_3$ & \cellcolor{red!20}$C_4$ & \cellcolor{red!20}$C_4$ & \cellcolor{red!20}$A_4$ &  \cellcolor{red!20}$A_3$ &  \cellcolor{red!20}$A_2$ &  \cellcolor{violet!30}$C_0$   & \cellcolor{violet!30}$C_0$   & \cellcolor{violet!30}$C_0$   & \cellcolor{violet!30}$C_0$    \\
        \hline
             11  & \cellcolor{green!20}$C_4$ & \cellcolor{green!20}$C_3$ & \cellcolor{green!20}$C_2$ & \cellcolor{green!20}$C_1$ & \cellcolor{yellow!20}$C_2$ & \cellcolor{red!20}$A_2$ &  \cellcolor{red!20}$A_3$ &  \cellcolor{red!20}$A_3$ &  \cellcolor{red!20}$A_3$ &  \cellcolor{red!20}$A_2$ &  \cellcolor{red!20}$A_2$ &  \cellcolor{violet!30}$C_0$   & \cellcolor{violet!30}$C_0$   & \cellcolor{violet!30}$C_0$   & \cellcolor{violet!30}$C_0$    \\
        \hline
             12  & \cellcolor{green!20}$C_4$ & \cellcolor{green!20}$C_3$ & \cellcolor{green!20}$C_2$ & \cellcolor{green!20}$C_1$ & \cellcolor{yellow!20}$C_2$ & \cellcolor{red!20}$A_2$ &  \cellcolor{red!20}$A_3$ &  \cellcolor{red!20}$A_4$ &  \cellcolor{red!20}$A_3$ &  \cellcolor{red!20}$A_2$ &  \cellcolor{red!20}$A_2$ &  \cellcolor{violet!30}$C_0$   & \cellcolor{violet!30}$C_0$   & \cellcolor{violet!30}$C_0$   & \cellcolor{violet!30}$C_0$    \\
        \hline
             13  & \cellcolor{green!20}$C_4$ & \cellcolor{green!20}$C_3$ & \cellcolor{green!20}$C_2$ & \cellcolor{green!20}$C_1$ & \cellcolor{violet!30}$C_0$   & \cellcolor{red!20}$A_2$ &  \cellcolor{red!20}$A_3$ &  \cellcolor{red!20}$A_3$ &  \cellcolor{red!20}$A_3$ &  \cellcolor{red!20}$A_2$ &  \cellcolor{violet!30}$C_0$   & \cellcolor{violet!30}$C_0$   & \cellcolor{violet!30}$C_0$   & \cellcolor{violet!30}$C_0$   & \cellcolor{violet!30}$C_0$    \\
        \hline
             14  & \cellcolor{green!20}$C_4$ & \cellcolor{green!20}$C_3$ & \cellcolor{green!20}$C_2$ & \cellcolor{green!20}$C_1$ & \cellcolor{violet!30}$C_0$   & \cellcolor{red!20}$A_2$ &  \cellcolor{red!20}$A_3$ &  \cellcolor{red!20}$A_4$ &  \cellcolor{red!20}$A_3$ &  \cellcolor{red!20}$A_2$ &  \cellcolor{violet!30}$C_0$   & \cellcolor{violet!30}$C_0$   & \cellcolor{violet!30}$C_0$   & \cellcolor{violet!30}$C_0$   & \cellcolor{violet!30}$C_0$    \\
        \hline
        \end{tabular}
        }

\vspace{3mm}
\centering
\begin{tabular}{|c|c|}
\hline
Notation & Symmetry type 
\\ 
 \hline\hline
$C_4$ & xyzt 
\\
 \hline
$C_3$ &  xyz* 
\\
 \hline
$C_2$ & xy** 
\\
 \hline
$C_1$ &  x*** 
\\
 \hline
$C_0\equiv A_1$ &  **** 
\\
 \hline
$A_4$ &  xyzt+\=x\=y\=z\=t 
\\
 \hline
$A_3$ &  xyz*+\=x\=y\=z* 
\\
 \hline
$A_2$ &  xy**+\=x\=y** 
\\
 \hline
\end{tabular}
    \caption{Symmetries for the $L=4$ case. Left: The fourteen weight cases of Fig.~\ref{fig:symmetries}(b) and the symmetries for different threshold intervals ($L=4$). This table summarizes how symmetry depends on the similarity threshold, $\alpha$ (different columns) in each case. The types of symmetry were obtained by numerically solving the system of ODEs and classifying the equilibrium in each case.
    The colors indicate the behavior of the associated stochastic model: green stands for consensus, yellow for polarization or consensus, red for polarization, and purple for persistent pluralism. Right: a list of the symmetry notations. Fig.~\ref{fig:symmetries}(a) visualizes each of these symmetries. The case of $L=5$ is presented in SI Section 6 and contains 1 intervals of $\alpha$ and 516 weight cases.}
    \label{tab:symmetries} 
\end{table}

In general, for a given set of weights, increasing $\alpha$ leads to a cascade of structural changes in the ODEs and the corresponding restructuring of their stable manifold. Two families of symmetry classes arise naturally in our opinion-dynamics model on the hypercube $\{0,1\}^L$, see Fig.~\ref{fig:symmetries} for $L=4$. The first family consists of cylinder sets, or subcubes, obtained by fixing a subset of higher-weight coordinates and allowing the remaining coordinates to vary freely. For example, the sets xyzt, xyz*, xy**, x***, and **** (for $L=4$) represent subcubes of dimensions 0 through 4. We denote these symmetries by $C_k$, where $k$ is the number of fixed coordinates. The second family consists of antipodal cylinder sets, obtained by taking a cylinder set and uniting it with its bitwise complement under the antipodal involution $x_i\mapsto 1-x_i$. For example, xyzt + \=x\=y\=z\=t, xyz*+\=x\=y\=z*, xy**+\=x\=y**  represent pairs of opposite subcubes. We denote these symmetries by $A_k$, where $k$ is again the number of fixed coordinates, see Table~\ref{tab:symmetries}(right) for a list of symmetries explaining the notations. Altogether, for $L=4$, we obtain eight fundamental symmetry classes: the cylinder sets $C_0$--$C_4$ and the antipodal cylinder sets $A_1$--$A_4$ (where $A_1$ is the same as $C_0$). For other issue numbers, see SI, table S2, for $L<4$, and SI Section 6 for $L=5$. We also note that for the case where all issues have the same weight ($w_i=1/L$), the system simplifies significantly, with only $L$ distinct $\alpha$ intervals, and no antipodal symmetries occurring, see SI Section 7.

The exact sequence of symmetry changes depends on the issue weights. Fig.~\ref{fig:symmetries} and Table~\ref{tab:symmetries} give the full classification of all the symmetries observed in the deterministic system with $L=4$.  All the sets of opinion weights  can be split into 14 different cases depending on the ordering of the sums of the weights subsets, see panel (b) of Fig.~\ref{fig:symmetries}.  Then, for each case, Table~\ref{tab:symmetries} lists the cascade of symmetries for all the $\alpha$ intervals, and the symmetries are shown in panel (a) of Fig.~\ref{fig:symmetries}. 


As seen from Table~ \ref{tab:symmetries}, in eight of the fourteen cases (cases 7-14), as $\alpha$ increases, the system goes through the symmetries depicted in the top row of Fig.~\ref{fig:symmetries}(a): no symmetry, followed by 8 pairs of strings that differ by only the last (least important) opinion (an example of this behavior is shown in panel (c6) of Fig.~\ref{fig:mod}), then four groups of four strings have the first two (most important) opinions in common, and then two groups of eight strings that have the first opinion in common. After this sequence, the behavior becomes more variable among the cases, and different types of symmetries appear, including symmetry $A_3$ that we saw in Fig.~\ref{fig:mod}(c). Of special importance is symmetry $C_0$, see Fig.~\ref{fig:symmetries}(a), bottom left: all the strings have exactly the same proportion, $1/2^L$. We will refer to this case as ``super-symmetric". As $\alpha$ increases, all the cases end up with super-symmetry (Table~\ref{tab:symmetries}). What sets this type of solution apart from the rest is the fact that it is the only asymptotically stable solution. That is, in the cases with super-symmetry, the stable manifold consists of a single point, without any ``neutral" directions (corresponding to the state of persistent pluralism).

While the exact trajectory of the deterministic model does not occur in the stochastic model, these symmetries do, as we already saw in Fig.~\ref{fig:mod}(c).

\begin{figure}
    \centering
    \includegraphics[width=0.7\linewidth]{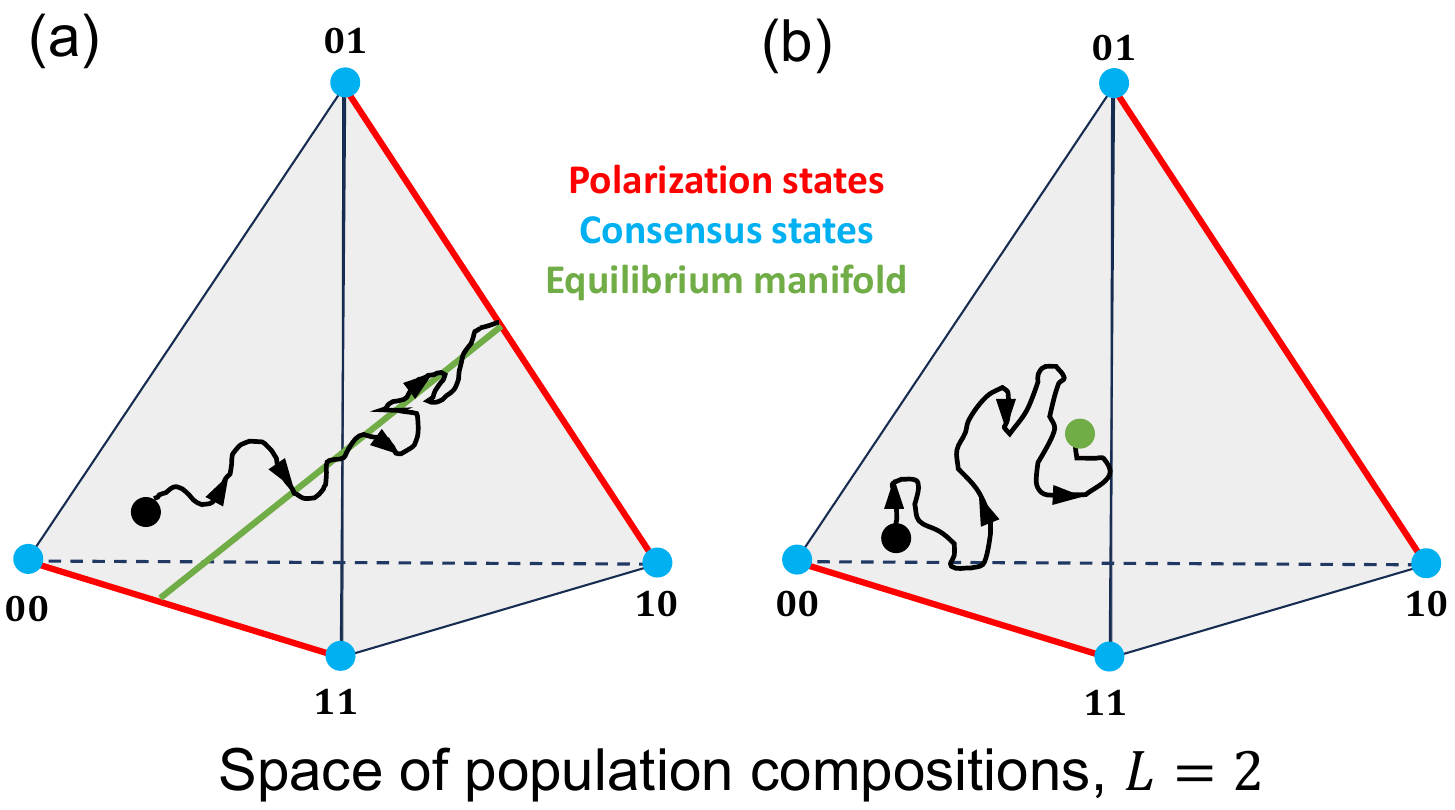}
    \caption{\textbf{The  equilibrium structure determined by symmetries explains the effects we observe.}  For $L=2$, the population composition is given by the proportions $(x_{00}, x_{01}, x_{10}, x_{11})$. The constraint $x_{00}+x_{01}+x_{10}+x_{11} = 1$ means one coordinate can be omitted, resulting in a tetrahedron (a 3D simplex) of points. Each vertex represents a population in a consensus state with a given opinion string, shown with blue dots. Lines between these represent populations with a combination of the two opinion strings at their endpoints. In particular, populations with only the opinion strings 00 and 11, or 01 and 10, are polarized and shown in red. A black trajectory indicates one possible path through this space the population composition could take under the opinion dynamics. For a large enough population size, this path will approach the manifold of equilibria shown in green. In the case on the left, this manifold connects to a polarized state, which will be reached relatively quickly. In contrast, the right panel features a case with only one attracting equilibrium. The dynamics will keep pulling the population back to this point and away from polarization or consensus, resulting in longer absorption times. Further, this case is significantly more likely to reach polarization than consensus, as the latter requires more opinion strings to vanish from the population than the former. This same principle applies to the larger $L$ cases. }
    \label{fig:explanation}
\end{figure}

\paragraph{Intuition and implications of the neutral drift manifold.} Using symmetry considerations, we can now understand  the effects observed, including the orders of magnitude larger absorption times for select ranges of the threshold value, the behavior of the probability of consensus vs polarization, and system destabilization by a small-weight issue. Stochastic opinion dynamics can be viewed as a walk in a multidimensional space (the dimensionality is given by the number of possible opinion strings minus one, $2^L-1$). For example,  If $L=2$, the composition of a population is given by four numbers $x_{00}, x_{01}, x_{10}$, and $x_{11}$, but since these sum to one we could determine $x_{11}$ from the first three, yielding motion in a 3D simplex, see figure \ref{fig:explanation}.  For large population sizes, the dynamics described by ODEs (\ref{master}) comprises the ``advection" part of the trajectories, with stochastic effects introducing fluctuations. Therefore, as time goes by, we expect the stochastic system to approach a steady state of the ODEs. 

There is however a fundamental difference between the long-term ODE and stochastic dynamics. The deterministic system converges to a stable manifold of solutions, which contains the symmetries discussed above. In  the example of \ref{fig:explanation}(a) where we assumed that $a<\alpha<a+b$ (see Fig. S3), the symmetry requires that $x_{01}=x_{01}$ and $x_{00}=x_{11}$; the corresponding manifold is marked in green. 
Outside the stable manifold, there is a deterministic force puling the solution back in (that is, $Re(\lambda)<0$ for the eigenvalues corresponding to the eigenvectors normal to the manifold). Inside the manifold, no such forces exist as the corresponding eigenvalues are zero. Now if we turn to the stochastic system, as mentioned above, the only true absorbing sets (under $r=0$) are either complete consensus or the existence of two opposing opinion strings (in arbitrary proportions); the set of absorbing states is marked by red in figure \ref{fig:explanation}. After entering the vicinity of the stable manifold of Eq. (\ref{master}), the stochastic system will tend to stay near it (because of the deterministic force pulling in in), but it continues to fluctuate, traversing the stable manifold by drift. As a result of this random (deterministically unbiased) motion, the stochastic system will eventually encounter one of its absorbing states as long as they are in or near the stable manifold. In the example of Fig. \ref{fig:explanation}(a), the polarization states $(x_{00}, x_{01}, x_{10}, x_{11}) = (0,0.5,0.5,0)$ and $(0.5,0,0,0.5)$ belong to the stable manifold and can be reached   by the stochastic system relatively quickly.

As the model parameters change, so does the symmetry, resulting in different outcomes being more likely. An interesting case is presented by the symmetry where all the strings have equal abundance (the super-symmetric solution), which in the example of figure \ref{fig:explanation}(b) is given by $x_{00}=x_{10}=x_{01}=x_{11}=\frac{1}{4}$. This stable manifold is the only such manifold that is comprised of a single point, and it is far from all the absorbing states of the stochastic system. Once the stochastic system is attracted to this point, it will proceed to fluctuate around it for a very long time (the state of persistent pluralism), as only an extremely large (and exceedingly rare) fluctuation will take it to absorption. It also follows that in such cases, polarization is the more likely outcome, as it requires all but two opinion strings to vanish from the population. In contrast, consensus requires one further opinion string to vanish, making it a rarer condition. Since super-symmetry happens for all the cases of weights for large values of $\alpha$, we can now see the emergence of extremely long absorption times for large $\alpha$ intervals in figure \ref{fig:effects}(c,d). \\

In Fig. \ref{fig:effects}(d), we also observe an extremely long convergence time for intermediate values of $\alpha$. The choice of weights in this example falls under case 13 of Table~\ref{tab:symmetries}, where we encounter super-symmetric solutions for the 5th $\alpha$ interval. For the threshold values in that interval the only stable solution of the ODE is a symmetric one, leading to an extremely long convergence times in the stochastic model. This is what is shown in panel (e) of Fig.~\ref{fig:effects}. The weight intervals for the given weights are shown by the vertical bar. Picking $\alpha$ from the thin 5th  interval (or from one of the last 5 intervals) results in a super-symmetric ODE solution and in a stochastic trajectory that oscillates around this point, i.e. the state of persistent pluralism. 

From Table~\ref{tab:symmetries} we learn that this happens in only two out of 14 possible weight cases, so it is a relatively rare phenomenon. For larger numbers of issues, there are more circumstance where one observes this behavior for intermediate $\alpha$-values. 

In Fig. \ref{fig:add-issue-and-symmetry} we observe how symmetry effects may lead to a counterintuitive destabilization of a polarized population by adding a very small-weight issue. This can also be explained by looking at solution symmetries. The original $L=3$ system quickly reached (near-)absorption (since a small amount of noise is present in this system, some small changes continue to occur). But after transitioning to an $L=4$ system, we find that for the given threshold $\alpha$, the only stable ODE solution is the super-symmetric one, and therefore there is an immediate strong deterministic force pulling the stochastic system away from the polarized state and toward the super-symmetric state. This is what is seen to the right of the black dashed line in figure \ref{fig:add-issue-and-symmetry}. The system will oscillate around this near-uniform opinion string distribution until an unlikely giant fluctuation will lead to system absorption. 

Finally, symmetries observed in the deterministic system help explain the consensus vs polarization behavior of the stochastic opinion dynamics. In Fig.~\ref{fig:effects}(a) and (c), we saw that small values of $\alpha$ are associated with consensus solutions and intermediate values of $\alpha$ with polarization solutions (while large $\alpha$'s lead to persistent pluralism). Fig.~\ref{fig:effects} shows more generally, how the long-term behavior of the stochastic system corresponds to the symmetry exhibited by the steady state of the ODEs. The green cells in Table \ref{tab:symmetries} denote convergence to consensus, yellow cells show the $\alpha$-regions where either consensus or polarization can be reached, red corresponds to polarization (where the solution of the stochastic system shows pairs of polar opposite strings changing in unison, like in figure \ref{fig:mod}(c1)), and purple is persistent pluralism (always associated with the super-symmetric ODE equilibrium). While there is no one-to-one correspondence between the rest of the colors and ODE symmetries, a closer look reveals several  patterns, consistent through all 14 weight cases: (a) polarization can be observed (yellow and red regions) if $w_3+w_4<\alpha<w_1+w_2$; (b) polarization is always observed (red regions) if $\min\{\max\{w_1,w_3+w_4\},w_2+w_3\}<\alpha<w_1+w_2$; (c) persistent pluralism is observed if $\alpha>w_1+w_2$. These rules allow constructing systems that maximize the region of polarization: for systems with $w_1+w_2\to 1$, that is, two important issues and two unimportant issues,  polarization may occur for almost any value of $\alpha$ (more precisely, for $0\approx w_3+w_4<\alpha<w_1+w_2\approx 1$), and for systems with $w_1\to 1$, i.e. one important issue and 3 unimportant issues, polarization will occur for almost any value of $\alpha$ (more precisely, for $0\approx w_2+w_3<\alpha<w_1+w_2\approx 1$). In other words, minimizing Shannon's entropy of the opinion weights leads to the highest chance of polarization in the population.

{\paragraph{Toward more realistic modeling.} We also examined a natural extension of the model, in which opinions on issues of different importance change at different rates (SI Section 10). Specifically, we assumed that opinions on highly weighted issues are updated more slowly than opinions on less important issues. While this modification leaves the qualitative structure of the model unchanged (that is, the same symmetry classes and long-term outcomes occur as in the baseline model), it substantially alters the transient dynamics. Opinion change proceeds hierarchically: disagreements on minor issues are resolved first, whereas the most important issues evolve on much longer time scales. Thus, issue importance primarily affects the speed and ordering of opinion change rather than the eventual collective outcome.

Another extension allows different individuals to assign different weights to the same issues (SI Section 11). We found that the principal phenomena identified in the homogeneous model remain robust: the system still exhibits transitions among consensus, polarization, and persistent pluralism as the similarity threshold varies. However, heterogeneity in issue importance enriches the dynamics and tends to enlarge the parameter regions where highly symmetric pluralistic states occur. As variation in issue weights across individuals increases, populations become more likely to sustain long-lived diversity of opinions rather than collapse to consensus or polarization.

\section*{Discussion}

We have investigated how opinions evolve in populations where individuals hold views on multiple issues of varying importance, and where social interactions can be either friendly (promoting agreement) or antagonistic (driving opinions apart). 


We find that small changes in the threshold parameter that determines whether interactions are attractive or repulsive, may lead to not only transitions between states (such as consensus vs polarization), but also to non-monotonic behavior (such as transitions from consensus to persistent pluralism to polarization and back to persistent pluralism). Perhaps most surprisingly, we discover that introducing even a single new issue of trivial importance can dramatically alter population dynamics, destabilizing previously stable consensus or polarization states, sometimes increasing convergence times by orders of magnitude. These counterintuitive effects arise from underlying symmetries in the dynamical system. For systems with up to $L=5$ issues, we provide a complete theoretical characterization explaining when and why these effects occur, classifying all possible weight configurations into a number of distinct cases, each exhibiting characteristic transitions as the similarity threshold varies. In particular,  we uncover the existence of two families of symmetry  in our opinion-dynamics model on the hypercube. (1) One family results in the population  split into several groups, such that within each group, the opinions on $k$ most important issues are shared, and the opinions on the rest of the issues are split equally among the group members.  (2) The second family forms groups with a more complex structure:  they consist of individuals that share the opinion on the first $k$ issues, together with individuals with polar opposite opinions on the same $k$ issues, while there rest $L-k$ less important opinions are split equally among the individuals. This latter symmetry types are strongly associated with the polarized solutions. The most symmetric solution where each string of opinions occurs equally often in the population leads to a quasi-equilibrium dynamics far from absorbing states, producing extremely long-lived states of persistent pluralism where all opinion strings coexist at roughly equal frequencies.

These results remain robust for two more realistic extensions of the model: when issue importance affects opinion-switching rates and when individuals assign different weights to the same issues. We found that the symmetries that characterize the basic model still inform the dynamics of the extended systems, although both extensions may modify the transient dynamics and can increase the prevalence of persistent pluralism. Our results complement earlier mechanisms that generate persistent coexistence of competing opinions. In classical majority-rule models, coexistence can emerge through the presence of contrarian agents or other forms of behavioral heterogeneity \cite{Galam2004contrarian,Galam2007role,Galam2023unanimity}. In our model,  long-lived diversity arises even in the absence of contrarians or stubborn minorities, as a consequence of the multidimensional structure of opinions and the interactions among issues with different weights.\\

Our findings have important implications for understanding polarization and consensus formation in real societies. The model suggests that two key factors determine whether a population reaches consensus, polarization, or persistent diversity: (i) the similarity threshold required for friendly interactions, and (ii) the relative importance assigned to different issues.

The role of the similarity threshold offers insights into why some societies may become polarized. When people are willing to engage constructively even with those holding somewhat different views (low $\alpha$), our model predicts consensus formation. However, when high similarity is required for positive interactions (a phenomenon increasingly observed in politically polarized societies where ``across the aisle'' friendships have declined) the model predicts either polarization into opposing camps or fragmentation into multiple factions. This aligns with empirical observations that declining tolerance for opinion differences correlates with increased political polarization \cite{stewart2021inequality}.\\

The finding that issue weights affect polarization risk provides practical insights. Our analysis shows that minimizing Shannon's entropy of issue weights, that is, concentrating importance on just a few issues while treating others as trivial, maximizes the parameter space leading to polarization. Conversely, distributing importance more evenly across issues reduces polarization risk. This suggests that societies may reduce polarization by broadening political discourse beyond a narrow set of highly salient issues and encouraging engagement with diverse policy domains. The introduction of new topics into public discourse, even seemingly minor ones, can shift the symmetry structure of opinion dynamics and potentially break polarization deadlocks.

However, our results also reveal that simply adding issues does not necessarily reduce polarization. The effect depends critically on both the weight assigned to new issues and the current state of the population. In some parameter regimes, introducing low-weight issues can paradoxically stabilize polarization rather than undermining it. 
Nevertheless, our general finding that persistent pluralism (rather than  polarization) becomes more likely as additional issues enter the discourse suggests that broadening political conversation may increase opinion diversity and reduce sharp divides.

To move populations from polarization toward greater diversity, our model suggests several strategies: (i) reducing the similarity threshold by fostering cross-cutting social connections and encouraging constructive engagement across opinion differences, (ii) introducing new issues to disrupt existing symmetries and create opportunities for coalition realignment, and (iii) reducing the disproportionate salience of a small number of divisive issues by broadening public discourse. These recommendations align with recent empirical work showing that exposure to politically diverse viewpoints can reduce polarization under appropriate conditions \cite{bail2018exposure}. \\

Our work extends existing opinion dynamics models in several ways. Classical bounded confidence models \cite{hegselmann2002opinion} demonstrated how limiting interactions to similar others can produce fragmentation, but typically considered only attractive forces and equal issue weights. More recent work has begun incorporating repulsive interactions \cite{chen2017deffuant,lanchier2024deffuant,huang2024breaking}, showing how antagonistic forces drive polarization. Our contribution lies in including the realistic assumption of issue weight heterogeneity and  systematically analyzing how heterogeneous issue weights interact with these opposing forces, revealing the rich symmetry structure underlying equilibria.  

The multidimensional nature of our model builds on Axelrod's cultural dissemination framework \cite{axelrod1997dissemination} and subsequent extensions \cite{baumann2021emergence,stamoulas2018convergence}, but our focus on heterogeneous weights combined with attraction-repulsion dynamics yields novel predictions. The symmetry-based analysis we develop provides a general framework that complements existing approaches to opinion dynamics \cite{castellano2009statistical,grabisch2020survey}.\\


Our model makes simplifying assumptions that warrant discussion. First, we assume well-mixed populations where any individual can interact with any other. In reality, social interactions occur on networks with specific topological features (clustering, degree heterogeneity, community structure) that might affect dynamics. Work on spatial voter models \cite{cox1995hybrid,durrett1994stochastic,durrett2006spatial} has shown that spatial structure alters convergence times and equilibrium outcomes. For instance, in spatial settings, clustering can maintain local consensus while global diversity persists. Extensions of our model to networks would likely reveal how network topology interacts with issue weight heterogeneity and attraction-repulsion forces. Previous work on opinion dynamics on networks \cite{Galam2002minority, altafini2013consensus, proskurnikov2016opinion, meng2018opinion,} suggests that community structure, degree distributions, and the presence of influential individuals all modulate consensus and polarization. Understanding how these network effects combine with the symmetry structures we identify remains an important open question.

Related to spatial structure, our model could be extended to coevolving networks where individuals can selectively sever ties with those holding dissimilar opinions and form new connections \cite{holme2006nonequilibrium,vazquez2008generic}. Such coevolution can lead to echo chambers and accelerated polarization, and understanding how issue weight heterogeneity affects these dynamics would be valuable.

Second, we treat issue weights as fixed, but in reality, the perceived importance of issues changes over time in response to events, media coverage, and shifting social norms. Allowing dynamic weights that respond to population opinion distributions or external shocks could reveal interesting feedback loops. Further, in this work we assume a fixed set of issues and investigate transitions that happen once an extra issue is added, but a systematic study of   emergence and fading of issues  (issue birth and death) would connect to work on innovation and cultural evolution \cite{hirshleifer2021moonshots}. 

Third, our model assumes binary opinions on each issue. While this simplifies analysis, real opinions often have more nuanced gradations. Extending to continuous opinion spaces \cite{hegselmann2002opinion,stamoulas2018convergence} while maintaining heterogeneous issue weights would be worthwhile, though the symmetry analysis would become considerably more complex.

Fourth, we assume homogeneous agents who all apply the same similarity thresholds. Introducing heterogeneity in thresholds, susceptibility to influence, or personal convictions could produce richer dynamics \cite{cheng2025multidimensional,parsegov2017novel}. Similarly, the presence of ``stubborn agents'' or opinion leaders with fixed views might alter outcomes \cite{Galam2007role, acemoglu2013opinion}.

Despite these limitations, our model provides interesting  theoretical insights into how heterogeneous issue importance and the balance between attraction and repulsion shape collective opinion dynamics. The symmetry-based framework we develop offers a powerful analytical tool for understanding multidimensional opinion formation, and the counterintuitive role of low-weight issues highlights the subtle ways that changes in discourse can trigger major shifts in collective outcomes. \\

This work demonstrates that opinion dynamics with heterogeneous issue weights exhibit surprisingly rich behavior arising from underlying symmetries in the system. Small changes in model parameters, including the introduction of seemingly trivial issues, can trigger dramatic shifts between consensus, polarization, and persistent pluralism. Our complete characterization for systems with up to five issues provides a foundation for understanding these phenomena, while incorporating issue weight heterogeneity in models of social dynamics. Moving forward, extensions to network structures or dynamic issue landscapes will deepen our understanding of how opinions evolve in complex societies. The tension between forces promoting consensus and those driving polarization is fundamental to social dynamics, and our mathematical model provides a tool for understanding when and how populations can overcome divisive dynamics to find common ground or maintain healthy pluralism.

\bibliographystyle{unsrt}
\bibliography{refs}



\end{document}